\documentstyle[11pt,IAUS215,twoside,epsf]{article}

\newcommand{\Sec}{\mathrm{s}}
\newcommand{\erg}{\mathrm{erg}}
\newcommand{\G}{\mathrm{G}}
\newcommand{\K}{\mathrm{K}}
\newcommand{\g}{\mathrm{g}}

\newcommand{\cm}{\mathrm{cm}}
\newcommand{\km}{\mathrm{km}}
\newcommand{\ms}{\mathrm{ms}}
\newcommand{\Junit}{\erg\,\Sec}
\newcommand{\junit}{\cm^2\,\Sec^{-1}}
\newcommand{\wunit}{\mathrm{rad}\,\Sec^{-1}}
\newcommand{\kms}{\km\,\Sec^{-1}}
\newcommand{\gcc}{\g\,\cm^{3}}

\newcommand{\Msun}{\mathrm{M}_{\odot}}
\newcommand{\NNmu}{N^2_{\mu}}
\newcommand{\NNT}{N^2_{T}}
\newcommand{\Br}{B_r}
\newcommand{\Bphi}{B_{\phi}}
\newcommand{\BB}{\Bphi\Br}

\newcommand{\Oc}{\Omega_{\mathrm{c}}}
\newcommand{\rhoc}{\rho_{\mathrm{c}}}
\newcommand{\oK}{\omega_{\mathrm{Kepler}}}

\newcommand{\powersep}{\times}
\newcommand{\Ep}[1]{10^{#1}}
\newcommand{\E}[1]{\powersep\Ep{#1}}

\def\edcomment#1{\iffalse\marginpar{\raggedright\sl#1\/}\else\relax\fi}
\reversemarginpar

\begin{document}
\vspace*{1cm}
\title{Presupernova Evolution of Rotating Massive Stars\\
and the Rotation Rate of Pulsars}
\author{A.\ Heger}
\affil{Department of Astronomy and Astrophysics, Enrico Fermi
  Institute, University of Chicago, 5640 S. Ellis Avenue, Chicago
  IL 60637, USA}
 \author{S.\ E.\ Woosley}
\affil{Department of Astronomy and Astrophysics, UCSC, Santa Cruz CA
  95064, USA}
\author{N.\ Langer}
\affil{Astronomical Institute, P.O. Box 80000,
NL-3508 TA Utrecht, The Netherlands}
\author{H.\ C.\ Spruit}
\affil{Max-Plank-Institut f\"ur Astrophysik, Karl-Schwarzschild Str. 1, 85740 Garching, Germany}

\begin{abstract}
Rotation in massive stars has been studied on the main sequence and
during helium burning for decades, but only recently have realistic
numerical simulations followed the transport of angular momentum that
occurs during more advanced stages of evolution.  The results affect
such interesting issues as whether rotation is important to the
explosion mechanism, whether supernovae are strong sources of
gravitational radiation, the star's nucleosynthesis, and the initial
rotation rate of neutron stars and black holes.  We find that when
only hydrodynamic instabilities (shear, Eddington-Sweet, etc.) are
included in the calculation, one obtains neutron stars spinning at close
to critical rotation at their surface -- or even formally in excess of
critical.  When recent estimates of magnetic torques (Spruit 2002) are
added, however, the evolved cores spin about an order of magnitude
slower.  This is still more angular momentum than observed in young
pulsars, but too slow for the collapsar model for gamma-ray bursts.
\end{abstract}

\section{Introduction}

Stars more massive than about $8-10\,\Msun$ end their life as
supernova, leaving a compact remnant - a neutron star or a black hole.
Generally speaking, the evolution of such a star follows a well
understood path of contraction to increasing central density and
temperature.  This path of contraction is interrupted by nuclear
fusion -- first hydrogen to helium, then helium to carbon and oxygen,
followed by carbon, neon, oxygen, and silicon burning, until finally a
core of iron is produced.  Each fuel burns first in the center of the
star, then in one or more shells.  In Table~1 we summarize the burning
stages and their durations for a $20\,\Msun$ star.  The time scale for
helium burning is about ten times shorter than that of hydrogen
burning, mostly because of the lower energy release per unit mass.
The time scale of the burning stages and contraction beyond central
helium-burning is greatly reduced by thermal neutrino losses that
carry away energy \textit{in situ}, instead of requiring that it be
transported to the stellar surface by diffusion or convection.  These
losses increase with temperature as roughly $T^9$ (see Woosley,
Heger \& Weaver for a more extended review).  When the star has built
up a large enough iron core, exceeding its effective Chandrasekhar
mass, it collapses to form a neutron star or a black hole.  A
supernova explosion may result (e.g., Colgate \& White 1966), or, in
rare cases, a gamma-ray burst (Woosley 1993; MacFadyen \& Woosley
1999; MacFadyen, Woosley, \& Heger 2001).

\begin{table}[b]
\caption{Nuclear burning stages in massive stars.  The table gives
burning stages, main products (ashes), typical temperatures, and time
scales for a $20\,\Msun$ star.}
\vspace{-\baselineskip}
\begin{center}
\setlength\tabcolsep{2ex}
\begin{tabular}{cccc}
\hline\noalign{\smallskip}
Fuel & Main &
$T$ & duration  \\
 & Product & ($10^9\,$K) & (yr) \\
\noalign{\smallskip}
\hline
\noalign{\smallskip}
H  & He     & 0.037 & $8.1\E6$ \\
He & O, C   & 0.19  & $1.2\E6$ \\
C  & Ne, Mg & 0.87  & $9.8\E2$ \\
Ne & O, Mg  & 1.6   & 0.60      \\
O  & Si, S  & 2.0   & 1.3       \\
Si & Fe     & 3.3   & 0.031     \\
\noalign{\smallskip}
\hline
\noalign{\smallskip}
\end{tabular}
\end{center}
\end{table}

Unevolved massive stars, however, are known to rotate rapidly at
several times 10\,\% of their break-up velocity at the surface.
Characteristic values of the equatorial rotation velocity on the main
sequence are around $\sim200\,\kms$ (Fukuda 1982; and other
contributions in these proceedings).  In this paper, we address what
happens to the rotation in the interior of these stars as they
evolve. In general, the ongoing contraction leads to faster rotation,
but transport processes (circulation, viscosity, magnetic fields,
etc.) lead to loss of angular momentum from the stellar core.  We are
particularly interested in the model predictions for the rotational
periods of the collapsed remnants left by the supernova.

\section{Rotating stars}

\subsection{Angular momentum transport}

On the main sequence, massive stars quickly settle into rigid rotation
(e.g., Zahn 1992; Talon et al.\ 1997, Meynet \& Maeder 2000; Maeder \&
Meynet 2000).  Contraction increases the rotational energy faster than
gravitational binding energy and would eventually make the core
unstable to triaxial deformations (Ostriker \& Bodenheimer
1973). Without the transport of angular momentum, for typical initial
rotation rates, this would happen during central helium burning, but
clearly the transport of angular momentum in the stellar interior
cannot be neglected.  Important processes for angular momentum
transport include hydrodynamic instabilities (convection, shear,
circulation, etc.; see, e.g., Endal \& Sofia 1978; Knobloch \& Spruit
1983; Heger, Langer, Woosley 2000; Tassoul 2000) and magnetic fields
(e.g., Spruit \& Phinney 1998; Spruit 2002).  The instabilities,
however, are inhibited by thermal, and, most importantly,
compositional stratification inside the star (e.g., Kippenhahn 1974,
Knobloch \& Spruit 1983; Maeder \& Meynet 1997; Maeder 2003).

\begin{figure}
\plotone{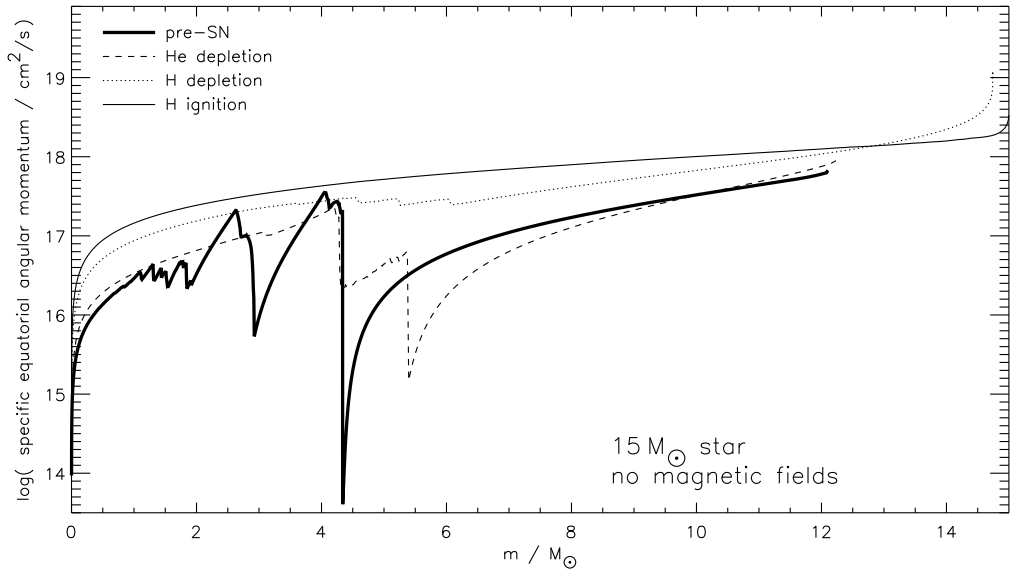}
\caption{Specific angular momentum, $j$, as a function of mass
coordinate, $m$ at different evolution stages in $15\,\Msun$ star
rotating with $200\,\kms$ on the zero-age main sequence (ZAMS;
hydrogen ignition).  The \textsl{thin sold line} gives $j$ on the ZAMS
at which the star is essentially in rigid rotation.  The
\textsl{dotted line} gives the angular momentum at central hydrogen
depletion ($1\,\%$ hydrogen left), the \textsl{dashed line} at central
helium depletion ($1\,\%$ helium left), and the \textsl{thick solid
line} the distribution at onset of core collapse.  The lines of later
evolution stages end at lower mass coordinate because of mass loss due
to stellar winds.  }
\end{figure}

Early calculations by Kippenhahn \& Thomas (1970); Kippenhahn,
Meyer-Hofmeister, \& Thomas (1970) and by Endal \& Sofia (1976) that
ignored angular momentum transport, or only included it in chemically
homogeneous regions, did indeed find that the star became secularly
unstable at the end of central helium burning or reached critical
rotation before central carbon burning.  Later models by Endal \&
Sofia (1978), that considered Eddington-Sweet circulation (Eddington
1924, 1929; Vogt 1925), shear instabilities (e.g., Zahn 1974, 1975;
Maeder 1997), GSF instability (Goldreich \& Schubert 1967; Fricke
1968), and Solberg-H{\o}iland instability (Wasiuty\'nski 1946) found
significant transport of angular momentum that allowed their stars to
go well beyond central carbon burning without reaching the limit of
secular instability.  Still the models rotated about three times
faster, after central helium burning, than the more recent models of
Heger, Langer \& Woosley (2000). This more recent study reduced the
efficiency of composition gradients in inhibiting instabilities and of
compositional mixing relative to angular momentum transport
(Pinsonneault et al. 1989; Chaboyer \& Zahn 1992) in order to
reproduce observed stellar surface abundance anomalies (e.g., Giess \&
Lambert 1992; Venn et al. 2002).  Calculations by the Geneva Group
also found, after central helium burning, rotation rates similar to
those of Heger et al. (2000; Maeder, private communication) and the
most recent calculations by Hirschi, Meynet, \& Maeder (2003) find
comparable rotation rates during oxygen shell burning, $0.16\,\wunit$
in $15\,\Msun$ star of initially $300\,\kms$ compared to
$0.12\,\wunit$ for a star with initially $200\,\kms$ in Heger et al.
Others have also studied massive star rotation, but concentrated on
the main sequence and helium burning (e.g., Urpin, Shalybkov, \&
Spruit 1996; Talon \& Zahn 1997; Denissenkov, Ivanova, \& Weiss 1999).

\begin{table}[!b]
\caption{Evolution of the internal stellar rotation for a $15\,\Msun$
star.  The first column gives the evolution stage (see explanation
below), the second column gives the central density at this evolution
stage, the 3$^{\it rd}$ column gives the specific angular momentum at a
mass coordinate of $1.7\,\Msun$, and the last three columns give the
radius, the angular velocity, and it's ratio to local Keplerian
angular velocity at this point.  For the neutron star (NS) we have put
the values in excess of critical rotation in brackets.}
\smallskip
\centering
\begin{tabular}{lrrrrr}
\hline\hline
Evolution                 & $\rhoc$   & $j$          & $r$         & $\omega$    & $\omega/\oK$ \\
stage                     & ($\gcc$)  & ($\junit$)   & ($\cm$)     & ($\wunit$)  & \\
\hline\noalign{\smallskip}
ZAMS$^a$         \dotfill & $5.8$     & $1.4\E{17}$ & $6.2\E{10}$ & $5.6\E{-5}$ & $0.058$ \\
TAMS$^b$         \dotfill & $11.7$    & $8.9\E{16}$ & $4.6\E{10}$ & $6.3\E{-4}$ & $0.041$ \\
He depletion$^c$ \dotfill & $2800$    & $3.7\E{16}$ & $8.0\E{9}$  & $8.6\E{-4}$ & $0.041$ \\
pre-SN$^d$       \dotfill & $6\E{9}$  & $2.8\E{16}$ & $2.8\E{8}$  & $0.56$      & $0.17$  \\
neutron star$^e$ \dotfill & $4\E{14}$ &$(2.5\E{16})$& $1.2\E{6}$  & $(3.4\E4)$  & $(3)$   \\
\hline
\end{tabular}
\smallskip\\
\small
\parbox{0.95\columnwidth}{
$^a$2\,\% hydrogen burnt; \\
$^b$1\,\% hydrogen left in the core; \\
$^c$1\,\% helium left in the core; \\
$^d$core collapse velocity reaches $1000\,\kms$; \\
$^e$for a $1.7\,\Msun$ core collapsing to a neutron star with moment
of inertia $I=1.44\E{45}\,$g$\,$cm$^2$ and assuming no angular
momentum loss. }
\end{table}

\subsection{Mixing}

Rotation also causes mixing even in regions that, without rotation,
are stably stratified.  For a review of mixing processes in stars, see
Pinsonneault (1997).  Observational evidence for this is the
surface enrichment of elements produced deep inside the star.  For
example, an enrichment of nitrogen with respect to oxygen and carbon,
or of helium, and the depletion of boron during central hydrogen
burning (e.g., Giess \& Lambert 1992; Venn et al. 2002; Fliegner,
Langer, \& Venn 1996).  The enrichment of helium above the convective
core during central hydrogen burning can also lead to an increase of the
luminosity similar to core overshooting that was often assumed in
non-rotating stellar models.

\begin{figure}[!t]
\plotone{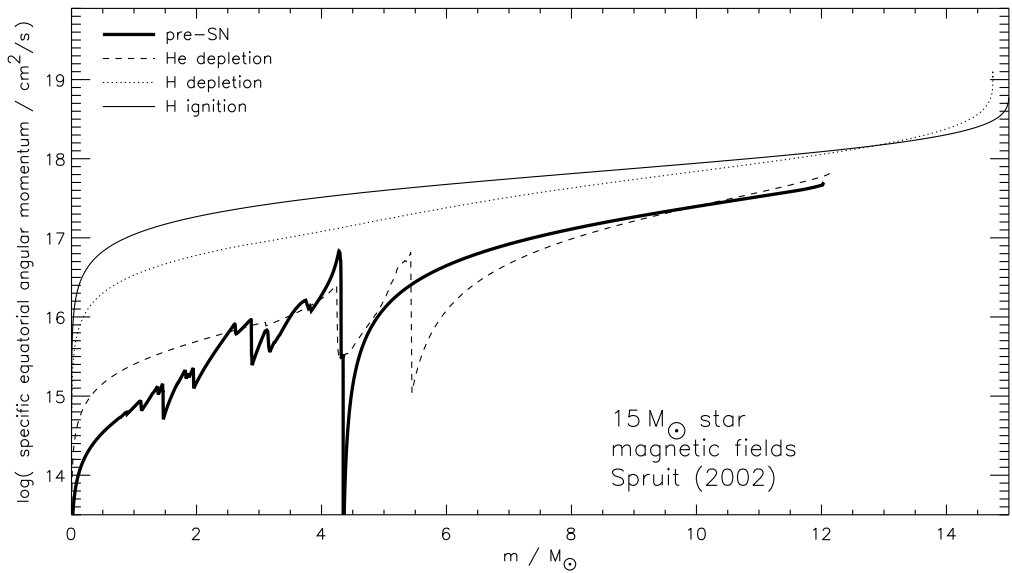}
\caption{Similar to Figure~1, but for a star that includes magnetic
torques according to Spruit (2002, 2003).}
\end{figure}

\begin{table}[!b]
\centering
\caption{Experimental periods (Marshall et al. 1998) and angular
momentum estimates for young pulsars.}
\smallskip
\begin{tabular}{lcccc}
\hline\hline\noalign{\smallskip}
                            & period & j	 & J          \\
\raisebox{1.5ex}[0pt]{pulsar}
                            & (ms)   & ($\junit$)  & ($\Junit$)   \\
\hline\noalign{\smallskip}
PSR J0537-6910 (N157B, LMC) &  16    & $2.0\E{14}$ & $5.67\E{47}$ \\
PSR B0531+21 (crab) \dotfill&  33    & $9.9\E{13}$ & $2.75\E{47}$ \\
PSR B0540-69 (LMC) \dotfill &  50    & $6.5\E{13}$ & $1.81\E{47}$ \\
PSR B1509-58  \dotfill      & 150    & $2.2\E{13}$ & $6.05\E{46}$ \\
\hline
\end{tabular}
\end{table}

During central helium burning, rotationally-induced mixing can prevent
the formation of layering in the convective core (Heger et al. 2002).
If the mixing above the convective core is fast enough, carbon and
oxygen can even be mixed outwards into hydrogen-rich regions and make
primary nitrogen, though more typically in stars of solar
metallicity, this is suppressed by a rise in entropy and the
composition barrier at the base of the hydrogen shell.  Additionally,
the ratio of circulation time-scale to evolutionary time is smaller
than during hydrogen burning, unless the core is spinning faster
($\tau_{\rm circ}\propto\tau_{\rm KH}(\Omega/\Omega_{\rm crit})^2$).
Note in particular, that hydrogen cannot be mixed down into the core,
since it will rapidly burn at the higher temperatures further down in
the star.  However, for stars of very low or zero metallicity, this
barrier is reduced and some primary nitrogen is made (e.g., Heger,
Woosley, \& Waters 2000a; Meynet \& Maeder 2002; Marigo, Chiosi, \&
Kudritzki 2003).  Unfortunately, the ``mixing events'' in such stars
are currently not well modeled and the amount of primary N$^{14}$
produced is not predicted very reliably.

Since the energy loss to neutrinos accelerates the later burning
stages, most rotationally-induced instabilities become too slow to mix
radiative regions (Heger et al. 2000), though shear instabilities may
still function at convective boundaries (see also Hirschi et al. 2003
in these proceedings).

\subsection{Evolution till core collapse and pulsar rotation}

In Figure~1 we give the evolution of the internal rotation profile for
a $15\,\Msun$ star and, in Table~2, numerical values for the specific
angular momentum, radius, and angular velocity at a mass coordinate of
$m=1.7\,\Msun$. This corresponds to the edge of the neutron star that
forms later.

\begin{table}[!b]
\caption{Magnetic Field and angular velocity evolution in a
$15\,\Msun$ star of solar composition and $200\,\kms$ equatorial
surface rotation on the ZAMS.  We give typical toroidal ($\Br$) and
radial ($\Bphi$) magnetic field strengths in the inner 1-$2\,\Msun$ if
radiative or outside the convective core for different evolution
stages (see below), and the central angular velocity.}
\smallskip
\centering
\begin{tabular}{lrrr}
\hline\hline
                          & $\Bphi$  & $\Br$    & $\Oc$       \\
\raisebox{1.5ex}[0pt]
{evolution stage}         & ($\G$)   & ($\G$)   & ($\wunit$)  \\
\hline\noalign{\smallskip}
MS$^a$           \dotfill & $5\E{3}$ & $1     $ & $5.7\E{-5}$ \\
TAMS$^b$         \dotfill & $1\E{4}$ & $2     $ & $2\E{-5}  $ \\
He ignition$^c$  \dotfill & $4\E{4}$ & $30    $ & $8\E{-5}  $ \\ 
He depletion$^d$ \dotfill & $1\E{4}$ & $2     $ & $6\E{-5}  $ \\
C ignition$^e$   \dotfill & $1\E{6}$ & $300   $ & $2.5\E{-4}$ \\
C depletion$^f$  \dotfill & $3\E{7}$ & $5\E{3}$ & $2\E{-3}  $ \\
O depletion$^g$  \dotfill & $5\E{7}$ & $7\E{3}$ & $4\E{-3}  $ \\
Si depletion$^h$ \dotfill & $3\E{8}$ & $2\E{5}$ & $7\E{-3}  $ \\
pre-SN$^i$       \dotfill & $5\E{9}$ & $\Ep{6}$ & $0.1      $ \\
\hline
\end{tabular}
\smallskip\\
\small
\parbox{0.6\columnwidth}{
$^a$20\,\% hydrogen burnt; \\
$^b$1\,\% hydrogen left in the core; \\
$^c$1\,\% helium burnt;  \\
$^d$1\,\% helium left in the core; \\
$^e$central temperature of $5\E8\,\K$; \\
$^f$central temperature of $1.2\E9\,\K$; \\
$^g$central oxygen mass fraction drops below $5\,\%$; \\
$^h$central Si mass fraction drops below $\Ep{-4}$; \\
$^i$infall velocity reaches $1000\,\kms$. }
\end{table}

During hydrogen burning, angular momentum loss by stellar winds and
the contraction of a core bounded by an expanding envelope leads to
angular momentum loss from the core.  As a gradient in mean molecular
weight develops, however, angular momentum gets ``trapped'' in the
core and differential rotation develops between the core and envelope.
By the end of central helium burning, rotation in the core is reduced
by another factor of two.  This is, in part, because the growing core
incorporates material of low angular momentum that had been slowed
down during the first dredge-up, and, in part, due to angular momentum
transport.  After central helium burning, hydrodynamic
rotationally-induced instabilities are not fast enough for significant
angular momentum transport (see above) and angular momentum is
essentially locally trapped in the cores of different composition.
Only convection is fast enough to redistribute angular momentum over
larger regions of the stars and the rigidly rotating convective shells
leave an imprint in the rotation profile that are each characterized
by a steep drop of the specific angular momentum at its base since,
typically, the inner radius of such a shells is much smaller than its
outer radius.

In the case of a $15\,\Msun$ star (solar metallicity) with an initial
rotation rate of $200\,\kms$ the core spins fast enough to form a
neutron star in excess of critical rotation (last line of Table~2) if
no further loss of angular momentum occurs.  Since rotation close to
breakup clearly is in contradiction to the rotation rates of young
pulsars (Table~3) some additional spin-down is necessary.  However,
this much rotation in collapsing massive star cores would be
preferable for the ``collapsar'' model for gamma-ray bursts (Woosley
1993).  For some time, the $r$-mode instability (e.g., Lindblom,
Tohline, \& Vallisneri 2001) seemed to be a promising explanation for
spin-down, so that both slow pulsars and collapsars could result from
rapidly rotating core, but recent evaluation (Arras et al. 2003) seem
to indicate that it may be not efficient enough (see also Woosley \&
Heger 2003 in these proceedings).  An alternative possibility for
removing this discrepancy may be angular momentum transport by
magnetic fields.

\section{Magnetic stars}

Spruit \& Phinney (1998) assumed that differential rotation would
torque an initially arbitrarily weak radial field component into a
toroidal field, that, when strong enough, would become unstable,
producing a new radial field that is wound up again.  With such strong
fields they obtained a star that was kept in rigid rotation until
about central carbon burning and estimated pulsar rotation rates at
birth $\sim100\,\Sec$ result.  If mass loss is taken into account, the
resulting spin would have been even lower by orders of magnitude.

\begin{table}[!b]
\centering
\caption{Pulsar rotation and angular momentum for different masses.
We assume initial solar composition and a ZAMS equatorial surface
rotation rate of $200\,$km$\,$s$^{-1}$.  In the 2$^{\it nd}$ column we
give the total angular momentum in the inner $1.7\,\Msun$ of the
stellar core.  Assuming no further loss of angular momentum and that a
neutron star with moment of inertia $I=1.44\E{45}\,$g$\,$cm$^2$ is
formed ($R=12\,$km, $M=1.4\,\Msun$, $I=0.36 M R^2$; Lattimer \&
Prakash 2001) the resulting (lower limits for the) pulsar periods are
given in the 3$^{\it rd}$ column.}
\smallskip
\begin{tabular}{lrrr}
\hline\hline\noalign{\smallskip}
stellar & J & period \\
mass & ($\Junit$) & ($\ms$) \\
\hline\noalign{\smallskip}
$15\,\Msun$ & $1.4\E{48}$ & 6.7 \\
$20\,\Msun$ & $1.8\E{48}$ & 5.0 \\
$25\,\Msun$ & $2.1\E{48}$ & 4.3 \\
\noalign{\smallskip}\hline
\end{tabular}
\end{table}

\begin{table}[!b]
\centering
\caption{Pulsar rotation rate dependence on dynamo model parameters.
For three different initial stellar masses of stars with solar
metallicity and ZAMS equatorial surface rotation rate of
$200\,$km$\,$s$^{-1}$ we give in the second column the resulting
neutron stars period (in ms) for a neutron star of assumed moment of
inertia $I=1.44\E{45}\,$g$\,$cm$^2$ of which we assume if forms form
the inner $1.7\,\Msun$ of the core of the star without further loss of
angular momentum.  The 3$^{\it rd}$ and 4$^{\it th}$ column give the
resulting period when the sensitivity to composition gradients is
lowered and raised by a factor 10.  The 5$^{\it th}$ and 6$^{\it
th}$ column give the result for a similar modification of thermal
buoyancy, and in the last two columns we give when the over-all
magnetic stress ($\sim\BB$) is modified by a factor 10 in both
directions.}
\medskip
\begin{tabular}{lrrrrrrr}
\hline\hline\noalign{\smallskip}
& 
& \multicolumn{2}{c}{\hrulefill\,$\NNmu$\hrulefill}
& \multicolumn{2}{c}{\hrulefill\,$\NNT$\hrulefill}
& \multicolumn{2}{c}{\hrulefill\,$\BB$\hrulefill}
\\
\raisebox{1.5ex}[0pt]{initial} 
& \raisebox{1.5ex}[0pt]{``std.''} 
& 0.1
&  10
& 0.1
&  10
& 0.1
&  10
\\
\raisebox{1.5ex}[0pt]{mass} 
& \multicolumn{7}{c}{\dotfill period ($\ms$) \dotfill} \\
\hline\noalign{\smallskip}
$15\,\Msun$ & 6.7 & 15 & 3.8 & 8.7 & 6.0 & 4.3 & 15 \\
$20\,\Msun$ & 5.0 & 11 & 2.3 & 5.3 & 4.4 & 2.8 & 9.6 \\
$25\,\Msun$ & 4.3 & 7.6 & 2.2 & 3.9 & 3.6 & 2.1 & 6.9 \\
\noalign{\smallskip}\hline
\end{tabular}
\end{table}

In a more recent approach, Spruit (2002) introduced a dynamo model
that considers stabilization of the stratification by thermal and
chemical stratification (see also Spruit 2003 in these proceedings).
It is assumed that the dynamo reaches a steady state field on a
time-scale short compared to that on which the structure and rotation
profile of the star evolve.  From this dynamo and its equilibrium
field an effective diffusivity for chemical mixing and an effective
viscosity for angular momentum transport can be derived (Spruit 2002).
We have implemented both as time-dependent diffusive processes 
(Heger, Woosley, \& Spruit 2003).
The results for a $15\,\Msun$ model with the same initial properties
as in Figure~1 and Table~2, but now including the dynamo process, is
given in Figure~2 and Table~4.  At the end of central hydrogen burning
the star is essentially still in solid body rotation -- the small
steps seen in Figure~1 between $4-6\,\Msun$, marking the composition
gradient left behind by central hydrogen burning, are not present --
and it rotates significantly slower.  By the end of central helium
burning, the core is slowed down even further.  At this point it
rotates at less than a tenth the speed of the non-magnetic star.
From here to core collapse the angular momentum in the inner
$2\,\Msun$ is reduced by an additional factor $\sim3$.  Despite
significant differential rotation, essentially no angular momentum is
transported across the edge of the helium core where entropy strongly
increases.  In Table~4 we also give the central angular velocity at
late evolution stages until core collapse along with the equilibrium
radial and toroidal magnetic fields.  This compilation demonstrates an
important property of of Spruit's model: the radial field component is
much weaker than the toroidal field component.  This is also one of
the significant differences with the model of Spruit \& Phinney
(1998).
From these stellar models we obtain neutron star rotation rates of
$\sim4-7\,\ms$ for $15-25\,\Msun$ stars assuming all the angular
momentum in the core were conserved to for a neutron star (Table~5)
and the remnants of more massive stars are rotating faster.  These
rotation rates are still faster than that of observed young pulsars
(Table~4) and additional spin-down mechanisms are still needed to
reduce the discrepancy (see Woosley \& Heger 2003 in
these proceedings).
In the derivation of the dynamo model by Spruit (2002) several
quantities of ``comparable size'' have been set equal, introducing
parameters ``of the order unity'' in the theory.  Most significant for
the angular momentum transport and the presence of effective barriers,
like the edge of the helium core mentioned above, are the influence of
stabilization of the stratification by thermal and compositional
buoyancy.  We have varied both, as well as the overall magnetic stress
from the dynamo by a factor 10 up and down (Table~6).  Even these
large variation only result in variations of the neutron star rotation
by about a factor $\sim2$.  The reason for the smallness of the effect
is that the stress resulting from the dynamo scales as the 6$^{\rm
th}$ power of the shear rate.

\section{Summary and conclusions}

Without the action of magnetic fields, hydrodynamic
rotationally-induced instabilities and convection alone (Figure~1) do
not transport angular momentum efficiently enough to avoid forming
pulsars rotating at break up (assuming that most of the angular
momentum is conserved during the collapse; Table~2).  When a dynamo
process like that of Spruit (2002) is included in the models (Table~4
and Figure~2), one obtains pulsar periods more than a factor 10
slower, about $4-7\,\ms$ (Table~5).  Though these periods may still
increase by $~20\,\%$ due to neutrino losses from the proto-neutron
star, they are still too fast compared even to the fastest-rotating
observed young pulsars (Table~3) and additional spin-down mechanisms
may still be needed in single stars (see Woosley \& Heger 2003 in
these proceedings).  However, the rotation rates found in the magnetic
stellar models would be too slow for most current gamma-ray burst
models which require rapidly rotating stellar cores.  However,
interacting close binary stars may lead to much faster or slower core
rotation, and, e.g., to pulsars of up to a second spin period (Langer et
al. 2003).

\smallskip

{\small
This research has been supported by the NSF (AST 02-06111), NASA
(NAG5-12036), and the DOE Program for Scientific Discovery through
Advanced Computing (SciDAC; DE-FC02-01ER41176).  AH is supported, in
part, by the Department of Energy under grant B341495 to the Center for
Astrophysical Thermonuclear Flashes at the University of Chicago, and
a Fermi Fellowship at the University of Chicago.
}


\end{document}